\documentclass[11pt,aps,tightenlines,nofootinbib,longbibliography,superscriptaddress]{revtex4-1}

\usepackage{graphicx,graphics,subfigure}
\usepackage{amsfonts,amsmath,amssymb,mathrsfs,bm,amsthm,nccmath}

\usepackage{array}
\usepackage{tabu}
\usepackage{dcolumn}
\usepackage{float}
\usepackage{url}
\usepackage{ragged2e}
\usepackage{mathtools} 
\usepackage{slashed}
\usepackage[usenames,dvipsnames]{color}
\usepackage{soul}
\usepackage{url}
\usepackage[colorlinks = true,
            linkcolor = blue,
            urlcolor  = blue,
            citecolor = blue,
            anchorcolor = blue]{hyperref}
\usepackage{physics}
\usepackage{makecell}
\usepackage{units}
\usepackage{upgreek}

\usepackage[total={6.5in,9in},top=1in,headsep=0.1in,headheight=1in]{geometry}
\usepackage[utf8]{inputenc}
\usepackage{charter}
\usepackage{fullpage}

\usepackage{lipsum}

\newcommand\snowmass{
\begin{center}
  \rule[-0.2in]{\hsize}{0.01in}\\
  \rule{\hsize}{0.01in}\\
  \vskip 0.1in
  Submitted to the Proceedings of the US Community Study\\
  on the Future of Particle Physics (Snowmass 2021)\\
  \rule{\hsize}{0.01in}\\
  \rule[+0.2in]{\hsize}{0.01in}\\[-2em]
\end{center}
}

\usepackage[firstpage=true]{background}
\backgroundsetup{contents={\parbox{6.5in}{\snowmass}}, scale=1,placement=top,opacity=1,color=black,position={3.25in,1.2in}}

\usepackage{fancyhdr}
\fancypagestyle{plain}{%
  \fancyhf{}%
  \fancyhead[C]{}
  \fancyfoot[C]{\thepage}
}

\begin{document}

\title{Snowmass2021 TF08 Whitepaper\\
Some open questions in axion theory}

\begin{abstract}
This white paper collects some open questions in several different aspects of axion theories. The questions are related to quantization of axion couplings, axion-magnetic monopole systems, axions in quantum gravity theory, axion string/domain wall systems, and thermal friction from axion couplings. They demonstrate many opportunities for axion theory, which call for more studies.  
\end{abstract}

\author{Prateek Agrawal}
\affiliation{Rudolf Peierls Centre for Theoretical Physics, University of Oxford, Parks Road, Oxford OX1 3PU, United Kingdom}
\author{Kim V. Berghaus}
\affiliation{C.N. Yang Institute for Theoretical Physics, Stony Brook University, NY 11794, USA}
\author{JiJi Fan}
\affiliation{Department of Physics, Brown University, Providence, RI, 02912, USA}
\affiliation{Brown Theoretical Physics Center, Brown University, Providence, RI, 02912, U.S.A}
\author{Anson Hook}
\affiliation{Maryland Center for Fundamental Physics, University of Maryland, College Park, MD 20742, USA}
\author{Gustavo Marques-Tavares}
\affiliation{Maryland Center for Fundamental Physics, University of Maryland, College Park, MD 20742, USA}
\author{Tom Rudelius}
\affiliation{Department of Physics, University of California, Berkeley}

 \maketitle
\flushbottom

\tableofcontents

\section{Introduction}
\label{sec:intro}

An axion is a periodic pseudo-scalar field, $a \cong a + 2\pi F$ with $F$ its fundamental period. It arises ubiquitously both from both top-down theories of quantum gravity~\cite{Svrcek:2006yi} and low-energy phenomenological models. The discrete shift symmetry acting on the field is its defining feature, and restricts the forms of its couplings. Axions enjoy a wide range of phenomenological and cosmological applications. The best known example is the QCD axion as a leading solution to the strong CP problem~\cite{Peccei:1977ur, Peccei:1977hh, Weinberg:1977ma,Wilczek:1977pj,Kim:1979if,Shifman:1979if,Zhitnitsky:1980tq,Dine:1981rt}. In addition, both the QCD axion and general axion-like particles (ALPs) serve as cold dark matter (DM) candidates through the misalignment mechanism~\cite{Preskill:1982cy,Dine:1982ah,Abbott:1982af}. 

Axion physics has been an active research field for more than four decades now, and interest in it has intensified recently. There has been continuous and tremendous progress in both the theoretical and experimental/observational probes. In this white paper, we will focus on multiple theoretical aspects of axion studies. We do not intend to provide a comprehensive review of all the theoretical developments, which is simply impossible to cover in a short white paper. There are already some very recent reviews covering part of the existing landscape of axion theories and models, e.g.~\cite{Choi:2020rgn, DiLuzio:2020wdo}. In addition, the community keeps developing a deeper understanding and making new discoveries about axion dynamics. Since the Snowmass process is a call to the future, this white paper collects some open questions appearing in active on-going axion research.

The questions collected here include how to measure and determine the quantization of axion couplings to massless gauge bosons, which is a defining feature of axions (Sec.~\ref{sec:quantizedcoupling}); develop a better understanding of dynamics in axion-magnetic monopole systems, in particular, axion potentials generated from either a real bath of monopoles (Sec.~\ref{sec:realmonopole}) or virtual monopole loops (Sec.~\ref{sec:virtualmonopole}); examine the axion properties in quantum gravity theories and sharpen the axion Weak Gravity Conjecture (Sec.~\ref{sec:quantum gravity}); calculate the QCD axion DM abundance from a string/domain wall system (Sec.~\ref{sec:dmabundance}); and improve the computation of the thermal friction induced by an axion coupling to a non-Abelian gauge group and explore further its applications to cosmology,  such as the thermal inflation scenario (Sec.~\ref{sec:thermalinflation}). 

This is by no means a complete list of interesting questions to be addressed in axion theory. We deliberately leave out a summary and conclusion section. If there has to be a summary, it will be: the open questions discussed in this white paper only serve as examples that the axion theory program is still a vibrant one with many opportunities to be explored, forty years after the axion model was first proposed. The study of axion theory should go hand in hand with the booming experimental efforts, and always requires further developments.


\section{Quantized axion couplings and how to measure them}
\label{sec:quantizedcoupling}

The discrete shift symmetry of the axion, $a \to a + 2\pi F$,  motivates the interpretation of the axion as a ``0-form" gauge field, in analogy to electromagnetic gauge invariance with the photon as a 1-form gauge field. The axion couplings to gauge bosons are topological and highly restrictive due to this fact. The coupling to a simple gauge group, e.g. $SU(N)$, can be written as
\begin{align}
    \mathcal{L} \supset 
    \frac{\mathcal{A} \alpha}{8\pi F} a\, G^{\mu\nu,a} \tilde{G}^a_{\mu\nu} \, ,
\end{align}
where $\alpha$ is the fine structure constant of $SU(N)$, $G^{\mu\nu,a}$ is the field strength and $\tilde{G}^a_{\mu\nu} = \frac{1}{2}\epsilon_{\mu\nu\rho\sigma} {G}^{\rho\sigma,a}$ is the dual field strength.
This operator resembles the Chern-Simons (CS) couplings in odd dimensions. In a similar fashion to CS theories, the anomaly coefficient $\mathcal{A}$ can be shown to be quantized. For constant axion field, the integral of the gauge field over spacetime (including a point at infinity) measures the winding number $\int \frac{\alpha}{8\pi} G \tilde {G} \in \mathbb Z$, so that in order for the theory to respect the gauged discrete shift symmetry of axion, $\mathcal{A}\in \mathbb{Z}$. It is a hallmark of CS couplings that they are not gauge invariant, but gauge invariant mod $2\pi$, and hence can appear with quantized coefficients in the action.
The quantization of the coupling may not always be apparent in a canonical basis due to mass mixing and kinetic mixing effects~\cite{Fraser:2019ojt}.

From a phenomenological point of view, one of the most important couplings is the axion coupling to photons,
\begin{align}
    \mathcal{L}_{a\gamma\gamma}
    &=
    -\frac14 g_{a\gamma\gamma} a F^{\mu\nu} \widetilde{F}_{\mu\nu}
    \, ,
\end{align}
where $F (\tilde{F})$ is the (dual) field strength of $U(1)$ electromagnetism. 
The mass of the axion $m_a$ and $g_{a\gamma\gamma}$ dictate much of axion phenomenology. Constraints are usually expressed in the $m_a - g_{a\gamma\gamma}$ plane, and arise from astrophysics, cosmology, black hole superradiance, helioscopes and terrestrial haloscopes for axion DM~\cite{Ouellet:2018beu,Salemi:2021gck,
    Asztalos2010,ADMX:2018gho,ADMX:2019uok,ADMX:2021abc,
    ADMX:2018ogs,Bartram:2021ysp,Crisosto:2019fcj,Lee:2020cfj,Jeong:2020cwz,CAPP:2020utb,Devlin:2021fpq,Grenet:2021vbb,HAYSTAC:2018rwy,HAYSTAC:2020kwv,McAllister:2017lkb,Alesini:2019ajt,Alesini:2020vny,CAST:2021add,DePanfilis,Gramolin:2020ict,Arza:2021rrm,Hagmann,Thomson:2019aht,ABRACADABRA,Liu:2018icu,Stern:2016bbw,Nagano:2019rbw,
    Ehret:2010mh,
    CAST:2007jps,CAST:2017uph,
    Betz:2013dza,
    OSQAR:2015qdv,
    DellaValle:2015xxa,
    SAPPHIRES:2021vkz,
	Xiao:2020pra,
	Keller:2021zbl,
    Chan:2021gjl,
    Wouters:2013hua,
    Marsh:2017yvc,
    Reynolds:2019uqt,
    Reynes:2021bpe,
    Reynes:2021bpe,
    Calore:2021hhn,
    Calore:2020tjw,
    Buen-Abad:2020zbd,
    Fermi-LAT:2016nkz,
    Meyer:2020vzy,
    Jacobsen:2022swa,
    HESS:2013udx,
    Ayala:2014pea,
    Wadekar:2021qae,
    Dessert:2022yqq,
    Li:2020pcn,
    Foster:2020pgt,
    Darling:2020uyo,
    Vinyoles2015,
    Jaeckel:2017tud,
    Payez:2014xsa,
    Dessert:2020lil,
    Blout:2000uc,
    Regis:2020fhw,
    Grin:2006aw}.
A thriving experimental program is underway to detect axions. In particular this is a field with a lot of synergy between theory and experiment, with a number of new experimental proposals on the horizon \cite{Lawson:2019brd,BRASS,
Liu:2021pei,DMRadio,Michimura:2019qxr,
Baryakhtar:2018doz,Beurthey:2020yuq,
Alesini:2017ifp,McAllister:2017lkb,
Schutte-Engel:2021bqm,Zhang:2021bpa,
Berlin:2020vrk, Meyer:2016wrm,Thorpe-Morgan:2020rwc,
Dekker:2021bos,Dolan:2021rya,Foster:2021ngm,Cadamuro:2011fd,Depta:2020wmr,
Ortiz:2020tgs, Shilon:2013xma,Ge:2020zww}, with several proposals already in the prototyping stage. We are now probing the most theoretically compelling parameter space of axion physics, and axions may well be the next discovery in fundamental physics.

The discovery of the axion will bring a wealth of information, precisely due to the nature of axion - photon couplings. We highlight two cases where this coupling can play a role in phenomenology. 

\subsection{QCD axion}
The QCD axion is a compelling candidate for new physics, resolving the strong CP problem and the DM puzzle at the same time. The quantization of the QCD axion coupling to photons is complicated by axion-pion mixing, but in a calculable way~\cite{GrillidiCortona:2015jxo}:
\begin{align}
    g_{a\gamma\gamma}
    &=
    \left(\frac{E}{N} - 1.92(4)\right) \frac{\alpha_{\rm em}}{2\pi f_a} \, ,
\end{align}
where the effective decay constant, $f_a$,\footnote{The fundamental period $F$ and the effective decay constant $f_a$ could be different, e.g., in multi-axion models such as Kim-Nilles-Peloso (KNP) alignment~\cite{Kim:2004rp}. While we distinguish them in this section, we will use them interchangeably in the following sections. } is defined via the axion-gluon coupling ,
\begin{align}
    \mathcal{L}_{a,{\rm QCD}}
    &= \frac{a}{8\pi f_a} \alpha_s G^a_{\mu\nu} \widetilde{G}^{\mu\nu,a}
    ,
\end{align}
which also sets the zero-temperature mass of the QCD axion~\cite{GrillidiCortona:2015jxo},
\begin{align}
    m_a
    &=
    5.70(7) \mu{\rm eV}
    \left(
    \frac{10^{12}\, {\rm GeV}}{f_a}
    \right).
\end{align}
The numbers $E$ and $N$ are the anomaly coefficients for the mixed anomaly. The ratio $E/N$ encodes valuable UV information. In Grand Unified Theories (GUT) with a simple unification group, the ratio $E/N$ is predicted to be $8/3$ with mild assumptions (standard indices of embedding of the SM gauge groups in the GUT). Thus, measuring this coupling could give us information about the far UV of the SM. 

The measurement of this coupling is an interesting challenge. To test simple GUTs, for example, even a few \% accuracy in $g_{a\gamma\gamma}$ may be sufficient. Axion haloscopes~\cite{ADMX:2021abc} will be able to measure the rate of axion-photon conversion to a very high precision after the initial discovery which will fix the axion mass. They will also be able to measure the coupling to gluons (equivalently $f_a$) at high precision under the hypothesis that the detected axion is the QCD axion. However, the rate measures a combination $g_{a\gamma\gamma}^2 \rho_{\rm DM}$ of the coupling and the local DM density. Unfortunately the local DM density is not known very precisely at the moment, and it will be challenging to pin it down to a few \%, even assuming that axion DM makes up the entire DM density. It will be interesting to investigate if it would be possible to identify a sub-component of DM with a more accurate local density prediction.

Another class of experiments that look for axion DM utilizes its coupling to gluons to measure a time-dependent oscillating electric dipole moment (EDM). Other observables which use derivative axion couplings do not help since these couplings are not quantized and undergo renormalization. The rate in these experiments involve the nuclear matrix elements which are only known to about 30\%~\cite{Pospelov:1999mv}. The EDM experiments and the photon experiments are not always operating in the same region of parameter space, but there is some region of overlap. If the axion exists in this overlap region, higher precision calculations of the matrix elements can break the degeneracy between the couplings and the local DM density. It will be interesting to develop other strategies that can measure and test the quantized couplings after the initial discovery.

\subsection{Hyperlight axions}
Topological defects, such as axion strings, are perfect candidates to give us direct access to the anomaly data of axion-photon couplings. For hyperlight axions, e.g. lighter than the Hubble parameter today $m_a < H_0$, axion strings formed in the early universe persist in the sky today. Such axions are a generic prediction of string compactifications, and even though these are not directly motivated from bottom-up problem solving perspective, they may well be the first axions we discover.

The axion-photon coupling in this case can be parametrized as (in the near-massless limit, the effects of mixing are unimportant)
\begin{align}
    \frac{\mathcal{A} \alpha_{\rm em}}{8\pi f_a} a F_{\mu\nu} \widetilde{F}^{\mu\nu} \, .
\end{align}
The anomaly coefficient $\mathcal{A}$ is quantized in units of the fundamental electric charge. 
\begin{align}
    \mathcal{A}
    &=
    \left(\frac{Q_{\rm fund}}{Q_{e}} \mathbb{Z} \right)^2 \, .
\end{align}
Linearly polarized light traveling in the background of an axion string undergoes a rotation of its plane of polarization by an angle $\Phi \sim \alpha_{\rm em}$. 
It was shown in~\cite{Agrawal:2020euj} that CMB polarization measurements are sensitive to such rotations, currently at the 1\% level, and expected to improve dramatically in next generation CMB experiments~\cite{,Jain:2021shf,Yin:2021kmx}. In particular, jumps of polarization across an axion string, measured in position space as opposed to spherical harmonic power spectra, are exactly quantized, and give us direct information about $\mathcal{A}$. Measuring $\mathcal{A}=1/9$, for instance, will tell us that the electron is not the fundamental unit of electromagnetic charge! It will be very interesting to develop these position space observables for upcoming CMB experiments and characterize their sensitivity to this quantized jump.


\section{Axion-magnetic monopole dynamics}

\subsection{Axion potential from a bath of magnetic monopoles}
\label{sec:realmonopole}

The Witten effect~\cite{Witten:1979ey} shows that if the electromagnetic $\theta_\gamma$ angle is non-zero, magnetic monopoles acquire electric charge proportional to $\theta_\gamma$. This can be easily seen looking at the static limit of Maxwell's equations for a non-zero $\theta_\gamma$,
\begin{equation}
    \begin{aligned} \label{eq:theta-em}
      \nabla \cdot \vec B = \rho_M \, , \\
      \nabla \cdot \left(\vec E - (\alpha/\pi) \theta_\gamma \vec B \right) = \rho_E \, ,
    \end{aligned}
\end{equation}
where $\rho_M$ ($\rho_E$) is the magnetic (electric) charge density, and $\theta_\gamma$ appears in the Lagrangian as
\begin{equation}
    \mathcal{L_\theta} = \frac{ \theta_\gamma \alpha}{\pi} \vec E \cdot \vec B \, . 
\end{equation}
From Eq.~\ref{eq:theta-em} we see that a monopole of magnetic charge $g$ also has electric charge $g e^2 \theta_\gamma /\pi$.

In theories with axions, this implies that magnetic monopoles develop an electric charge in an axion background, due to the coupling to photons, which for simplicity we will take to be
\begin{equation}
    \mathcal{L}_a = \frac{a}{f_a} \frac{\alpha}{ \pi} \vec E \cdot \vec B \, .
\end{equation}
The electric charge leads to an increase of the monopole self-energy, and thus, near the monopole the axion is pushed towards vanishing $(a/f_a + \theta_\gamma)$. This shows that there is a potential for the axion in the background of a monopole with a minimum corresponding to zero $\theta_\gamma$, similar to how the QCD instantons generate a potential for the axion with a minimum corresponding to vanishing strong CP phase $\bar \theta$, a combination of the QCD $\theta$ angle and the phase of the determinant of the Yukawa couplings. This was first pointed out in Ref.~\cite{Fischler:1983sc} (referred to as FP for the remainder of this section), where the form of the potential was also calculated under some approximations. Since there are a number of interesting phenomenological consequences for this effect (e.g.~\cite{Kawasaki:2015lpf,Nomura:2015xil}), it is important to revisit this topic and clarify whether important effects were included and if the approximate form of the potential is valid in general models. In what follows we point out a few important questions related to the calculation of the axion potential arising from the presence of monopoles.

Following the original analysis in FP, we can calculate the axion ground-state energy in the background of a monopole of minimal charge $g = 1/(2e)$ sitting at the origin, by minimizing the potential energy (assuming no other potential for the axion)
\begin{equation}
    V = \int d^3 r \left[ \frac{1}{2} (\nabla a)^2 + \frac{1}{2} \vec E^2 \right] = \int d^3r \left[ \frac{1}{2} (\nabla a)^2 + \frac{e^2}{128 \pi^4 r^4} \frac{a^2}{f_a^2} \right] \, .
\end{equation}
One can show that, assuming spherical symmetry, the configuration that minimizes energy with $a/f_a \rightarrow \theta_0$ ($\theta_0$ is a constant background value) as $r\rightarrow \infty$ is given by
\begin{equation}
    a(r) = f_a \theta_0 e^{-r_0/r} \, , \qquad r_0 = \frac{e}{8 \pi^2 f_a} \, ,
\end{equation}
and the minimum potential energy is
\begin{equation} \label{eq:preskill-potential}
    V_0 = \frac{e f_a \theta_0^2}{4 \pi} \, .
\end{equation}

Note that this calculation was done in the classical limit and also using the effective theory that is expected to be valid only at energies below $f_a$. However, one can easily see that for $r\sim r_0$,
\begin{equation}
    |\nabla a|^2 \sim \frac{16 \pi^3 f_a^4}{\alpha} \gg f_a^4 \, ,
\end{equation}
which implies that there might be non-negligible deviations from this result due to dynamics associated with the radial mode associated with the axion which was not included in the calculation. Understanding the effects from the radial mode in the axion-monopole dynamics is an important question that has not been yet addressed.

The other aspect of the calculation that could be improved is related to finite size effects, since the previous computation was only valid in the limit of a point-like monopole. As is discussed in FP, there are two important deviations from the field configuration associated with a point like monopole. The first is that the monopole itself is not expected to be point-like, and if the monopole radius, $r_c$ is larger than $r_0$, there should be large corrections to the analysis. In FP, the authors approximated this effect by changing the potential in Eq.~\ref{eq:preskill-potential} to the electrostatic self-energy of a charged sphere of charge $e \theta_0/2 \pi$ and radius $r_c$. A more rigorous analysis, including an analysis of potential impacts of the expected large gradients of the axion field in the monopole configuration, would be important to determine whether and in which regime this is a reasonable approximation.

The second important finite size effect is related to the presence of light charged states. It is well known that if there are massless charged fermions, the $\theta_\gamma$ term becomes unphysical, and so the axion potential generated by the monopole should also vanish in such limit. In FP, this effect was taken into account using the fact that electric fields would be screened over distances shorter than the inverse of the fermion mass, $m_\psi^{-1}$. Their conclusion was that in this case the potential in Eq.~\ref{eq:preskill-potential}, would be approximated as the electrostatics self-energy of a charged sphere of charge $e \theta_0/2 \pi$ and radius $m_\psi^{-1}$. A more robust derivation of this result would be important, since this can significantly weaken the effects of the potential generated by the monopole. Another aspect that was not discussed is screening due to the presence of a plasma of charged states. While for magnetic monopoles of a dark $U(1)$, it is not necessary that there are light charged states, in most applications to cosmology the dark $U(1)$ must eventually be broken so that the monopoles annihilate and don't lead to strong cosmological constraints. This implies that there must be some light states (e.g. the dark Higgs that would break the symmetry) which are charged. Therefore, it is important to determine how the monopole-generated axion potential gets modified in the presence of a bath of relativistic charged particles which would cause screening of electric fields at large distances.

The interactions between axions and magnetic monopoles have many interesting phenomenological consequences. In order to fully explore the cosmological implications of this dynamics, it will be important to make progress on the open questions presented in this section related to the axion potential generated by the presence of magnetic monopoles.

\subsection{Axion potential from virtual magnetic-monopole loops }
\label{sec:virtualmonopole}

The last section focuses on the axion potential from a real bath of magnetic monopoles, which could be generated via the Kibble-Zurek mechanism during a phase transition in the early universe~\cite{Kibble:1976sj, Zurek:1985qw}. In this section, we will discuss the axion potential from virtual magnetic-monopole loops. 

It was postulated in the Completeness Hypothesis that any UV completion of an interacting Abelian gauge theory with quantized charges contains magnetic monopoles~\cite{Polchinski:2003bq}. Furthermore, if there exists an axion coupling to the Abelian gauge field through $a F \tilde{F}/f_a$ coupling, a whole tower of excited dyonic states with both electric and magnetic charges should be present in the spectrum. As reviewed in the previous section, through the Witten effect~\cite{Witten:1979ey}, the monopole obtains an effective electric charge in the presence of an axion, which is proportional to the background value of $a/f_a$. When $a/f_a \to a/f_a + 2\pi$, a magnetic monopole becomes a dyon with one unit of electric charge. In fact, to preserve the periodicity of $a$, there must be a whole tower of dyons characterized by an integer quantum number $n$ with $n$ being the dyon's electric charge when the axion is absent. Turning on the axion background, the dyonic state with electric charge $n$ becomes the next state with $n+1$, when $a/f_a \to a/f_a + 2\pi$.

It was recently pointed out that virtual dyon loops could also lead to an effective potential for the axion, in the low-energy effective Abelian gauge theory. This effect exists and could be computed even when we have no knowledge of the UV completion of the Abelian theory. It is a consequence of the Witten effect. The dyon mass (energy) obtains a contribution from a volume integration of its electric field squared, which is determined by its electric charge squared. In the axion-dyon system, since the electric charges of the dyons are shifted by the axion field, their masses also become functions of the axion as well. We could integrate out these heavy dyonic states, sum over the resulting Coleman-Weinberg-type potentials, and obtain an effective potential for the axion~\cite{Fan:2021ntg}. Another approach is to treat the collective dyonic coordinate as another compact spatial direction in which the monopole could propagate. Then we could integrate over all the closed trajectories with different winding numbers, which yields the same result. Analogous to the dilute instanton approximation, the computations in Ref.~\cite{Fan:2021ntg} involving dyons assume that the effective axion potential is dominated by a single dyon loop and ignore the long-range Coulomb interactions between the dyons. The validity of the assumption needs to be checked. 

This new virtual dyon effect could have interesting phenomenological implications, which calls for further studies. In the simplest model of an Abelian hidden sector with an axion, this effect could be the dominant contribution to the axion mass and determines its cosmic abundance in a large region of the parameter space~\cite{Fan:2021ntg}.   

One important open question, similar to one of the finite size effects discussed in the previous section, is: how does the axion potential induced by dyonic loops change, in the presence of charged fermions?  As is familiar from the non-Abelian instanton physics, the dependence on the $\theta_\gamma$ angle should disappear in the limit when the charged fermions become massless and restore a chiral symmetry. It still remains to be shown how to incorporate the fermions in the computation of dyonic loops, which could reproduce this special limit. One crucial first step is to understand the interactions between magnetic monopoles/dyons and massless fermions. While this had been studied since the early 1980's~\cite{Callan:1982ac, Callan:1982ah, Callan:1982au, Rubakov:1982fp, Callan:1983ed}, there still remain many intriguing confusions, and it continues to be an active research topic~\cite{Grossman:1983yf, Nair:1983ps,Yamagishi:1982wp,Yamagishi:1983ua, Yamagishi:1984zu, Panagopoulos:1984ws, Sen:1984qe, Sen:1984kf,Balachandran:1983sw,Polchinski:1984uw, Preskill:1984gd,Kitano:2021pwt,Csaki:2021ozp,Brennan:2021ucy, Brennan:2021ewu}. The computation to use the fermion-dyon interaction to understand its impact on the axion potential in the axion-dyon system is not yet available. Solving this could address related interesting phenomenological questions such as whether there exists a minimum mass of the axion coupling to our electromagnetism, even without non-Abelian instantons.  \\

Lastly it will be interesting to explore axion models with an electromagnetic duality invariance (equivalently, $SL(2,Z)$ invariance), and thus a complexified coupling $\tau = \frac{\theta_\gamma}{2\pi} + i \frac{2\pi}{e^2}$~\cite{Montonen:1977sn}. In this case, an axion comes together with a saxion. The $SL(2,Z)$ symmetry strongly constrains the form of the scalar potential for both axion and saxion. It would be interesting to clarify the relation between the non-perturbative constraint from the symmetry and perturbative calculations (e.g., in a theory with electromagnetic duality, one may think that the electron/magnetic monopole loop contributes to the saxion potential in the perturbation calculation, and thus the axion potential as well, with a result that does not appear to respect the axion shift symmetry). Furthermore, could it be tested experimentally through axion couplings whether there is an electromagnetic duality?

\section{Axions in quantum gravity }
\label{sec:quantum gravity}

Quantum gravity is famously difficult to probe experimentally,  but axions are one of the most likely candidates for bridging the gap between theory and experiment. Axions are ubiquitous in well-understood string compactifications \cite{Svrcek:2006yi, Arvanitaki:2009fg}, and the crucial role played by axions in ensuring the absence of global symmetries in quantum gravity suggests that this will continue to hold more generally in the quantum gravity landscape \cite{Heidenreich:2020pkc}.

One challenge facing axions in the quantum gravity landscape is the ``axion quality problem'' \cite{Barr:1992qq, kamionkowski:1992mf, holman:1992us, Ghigna:1992iv}: quantum gravitational effects from Planck-scale physics can move the minimum of the QCD axion potential away from $\bar{\theta} = 0$, thereby spoiling the axion solution to the strong-CP problem. This problem can be avoided if Planck-scale contributions to the axion potential are negligible or if they are aligned so that the minimum of the potential remains very close to $\bar{\theta} = 0$. A number of approaches to the axion quality problem have been proposed \cite{Witten:1984dg, Randall:1992ut, Dias:2014osa, Lillard:2018fdt, Cox:2019rro, Hook:2019qoh, Ardu:2020qmo, Heidenreich:2020pkc, Darme:2021cxx}, but it is an open question within quantum gravity to determine whether or not these approaches can be successfully implemented \cite{Alvey:2020nyh, Yin:2020dfn} or if the quality problem itself is even of concern. See \cite{Hook:2018dlk} for a brief review.

Axions also play a starring role in discussions of large-field inflation. The periodic shift symmetry $a \rightarrow a + 2 \pi f_a$ naturally protects axions from Planck-suppressed operators, which generically spoil large-field inflationary models. Axions acquire a periodic potential from instantons of the form $V(a) = \Lambda^4 \sin(a/f_a)$, which for $f_a > M_{\textrm{Pl}}$ leads to a large-field model of inflation called natural inflation \cite{Freese:1990rb}, which can be distinguished from other models of inflation by the spectral tilt and tensor-to-scalar ratio of the primordial power spectrum.

However, quantum gravity appears to censor such natural inflation models. Within string theory, axion decay constants are constrained in all known examples to satisfy $f_a < M_{\textrm{Pl}}$ \cite{Banks:2003sx}, and $f_a < M_{\textrm{Pl}}$ is a generic prediction of the axion version of the ``Weak Gravity Conjecture'' \cite{ArkaniHamed:2006dz}. Models of natural inflation involving multiple axions \cite{Liddle:1998jc, Dimopoulos:2005ac, Kim:2004rp} are similarly constrained by the Weak Gravity Conjecture \cite{Rudelius:2014wla, Rudelius:2015xta, Montero:2015ofa, Heidenreich:2015wga, Brown:2015iha, Heidenreich:2019bjd}. Although loopholes to these constraints exist \cite{delaFuente:2014aca,Rudelius:2015xta,Brown:2015iha, Bachlechner:2015qja, Hebecker:2015rya}, it is unclear whether or not these loopholes can be successfully threaded in a UV complete theory of quantum gravity \cite{Brown:2015lia}. Indeed, surveys of axion landscapes in string theory have so far found them to be barren of large-field natural inflation \cite{Rudelius:2014wla, Conlon:2016aea, Long:2016jvd}. Determining whether or not such models are compatible with quantum gravity, either through an explicit example in string theory or a strict no-go theorem, remains a significant open question.

In addition to its application to large-field inflation, the axion Weak Gravity Conjecture also constrains the QCD axion decay constant \cite{Heidenreich:2016jrl} and models of early dark energy \cite{Rudelius:2022gyu}. Yet, the conjecture itself is quite mysterious: while the ordinary Weak Gravity Conjecture is motivated by demanding consistent black hole decay, the physical motivation underlying the axion Weak Gravity Conjecture is unclear, and the primary evidence for the conjecture comes simply from examples in string theory and from connections to the ordinary Weak Gravity Conjecture. The precise version of the axion Weak Gravity Conjecture is also an open question: the conjecture implies a bound $f_a S < c M_{\text{Pl}}$, where $S$ is the instanton action and $c$ is an order-one coefficient, but the precise value of the coefficient $c$ is still debated \cite{Heidenreich:2015nta, Hebecker:2016dsw, Hebecker:2018ofv, Andriolo:2020lul}. Answering this question likely requires an improved understanding of gravitational instantons and Euclidean wormholes in theories with axion fields, a topic which has been studied for several decades \cite{Giddings:1987cg, Coleman:1988cy,  Giddings:1988cx, Tamvakis:1989aq, Rubakov:1996cn,  Rey:1998yx, Gutperle:2002km, Bergshoeff:2004pg, Arkani-Hamed:2007cpn, Heidenreich:2015nta, Hebecker:2016dsw} but remains mysterious \cite{Hebecker:2018ofv}. Axion Weak Gravity Conjecture constraints on cosmology depend crucially on the value of the coefficient $c$, so determining which value (if any) is the one is an urgent task for the study of axions in quantum gravity.

Aside from instantons, axion potentials in string theory can also receive contributions from fluxes, which lead to models of axion monodromy inflation \cite{Silverstein:2008sg, McAllister:2008hb} in which the axion winds many times around its fundamental domain $a \in [0, 2 \pi f_a)$. In this setting, instanton contributions can lead to small, sinusoidal wiggles in the axion potential, which may lead to the observable phenomenon of resonant non-gaussianity in the CMB \cite{Flauger:2010ja, Flauger:2014ana}. Axion monodromy also plays a significant role in the relaxion solution to the hierarchy problem \cite{Graham:2015cka}.

Although quantum gravity does not impose significant constraints on axion monodromy or relaxion models through the Weak Gravity Conjecture \cite{Hebecker:2015zss, Ibanez:2015fcv}, backreaction is a more significant worry \cite{Valenzuela:2016yny, Buratti:2018xjt, McAllister:2016vzi}. A refined version of the Swampland Distance Conjecture \cite{Ooguri:2006in, Klaewer:2016kiy} suggests that super-Planckian traversals of a monodromy axion should be accompanied by towers of light states, which may backreact on the low-energy effective field theory \cite{Baume:2016psm} and/or the internal geometry of a string compactification. Whether or not these backreaction effects can be controlled in a UV complete theory of quantum gravity is a controversial question meriting further analysis.

There has also been significant work devoted to understanding universal features of axion strings in quantum gravity. The Weak Gravity Conjecture places a bound on the tension of an axion string of the form $T \leq c' f_a M_{\textrm{Pl}}$, but once again the precise value of the coefficient $c'$ remains an important open question \cite{Hebecker:2017wsu} with applications to black hole physics \cite{Bowick:1988xh,  Hebecker:2017uix}. In theories with $a F \tilde{F}$ couplings, the excitations of the axion string will satisfy the Weak Gravity Conjecture for the gauge field provided the axion string satisfies its version of the Weak Gravity Conjecture, leading to a mixing between of different versions of the Weak Gravity Conjectures \cite{Heidenreich:2021yda, Kaya:2022edp} which mirrors the mixing of background gauge fields into a higher-group symmetry \cite{Brennan:2020ehu}. These calculations generally lack precise order-one coefficients, however, so an improved understanding of axion strings could lead to a sharpening of the Weak Gravity Conjecture and more precise constraints for quantum gravity on low-energy effective field theory.

Axion strings have also served as a useful setting for studying the prospects of trans-Planckian scalar field excursions (as required in large-field inflation) \cite{Dolan:2017vmn, Draper:2019zbb}, and they play an important role in understanding and classifying infinite-distance limits of scalar field moduli space \cite{Lee:2019xtm, Lee:2019wij, Lanza:2021udy}. It has been conjectured that, in fact, \emph{every} infinite-distance limit of moduli space corresponds to an RG flow UV endpoint of an axionic string  \cite{Lanza:2021udy}, and proving this conjecture remains an interesting open problem.

For more information on potential quantum gravity constraints on axions, see the reviews \cite{Brennan:2017rbf, Hebecker:2018ofv, Palti:2019pca, vanBeest:2021lhn, Grana:2021zvf, Harlow:2022gzl}.

\section{QCD Axion DM from strings/domain walls }
\label{sec:dmabundance}
Axion string and domain wall dynamics in the early universe is extremely exciting.
In the context of axion DM, string and domain wall dynamics comes to the fore.  The reason for this is that there is a popular mechanism for producing QCD axion DM called the post-inflationary PQ symmetry breaking scenario~\cite{Preskill:1982cy,Abbott:1982af,Dine:1982ah}.  In this approach towards axion DM, the universe is assumed to reheat to a temperature above the PQ phase transition.  In this high temperature phase, the PQ symmetry is restored.  As the universe cools down, it undergoes a phase transition into the broken phase.  When this phase transition occurs, the axion can take a field value anywhere between $0$ and $2 \pi f_a$.  Different Hubble patches choose different values and by complete accident can form topologically non-trivial objects such as strings.  This mechanism of producing strings is called the Kibble-Zurek mechanism~\cite{Kibble:1976sj, Zurek:1985qw} and has been observed to occur in many condensed matter systems.  The initial number density and length distribution of strings is highly dependent on the details of the phase transition.  However, it is believed that the axion string network approaches a scaling solution regardless of its initial dynamics.  Eventually at the QCD phase transition, the axion obtains a potential and domain walls form.  These domain walls will stretch between axion strings and pull them together, destroying the string network.

This entire process produces many cold axions that can play the role of DM.  The sources of DM are roughly divided into three categories : misalignment, strings, and domain walls.  Firstly, away from the string the axion expectation value is somewhere between $0$ and $2 \pi f_a$.  After the axion obtains a potential, this initial condition gives some number of axions from the standard misalignement mechansm.  Secondly, at the scaling solution, the axion string network is constantly radiating axions that can contribute significantly to the DM abundance.  Finally, the collapse of the string network via the production of domain walls converts all of the energy in the string network into axions.  Together these three sources of axions give a prediction for the abundance of DM.  Fixing the abundance of DM to its observed value predicts a value for the mass of the QCD axion.

This mechanism for producing QCD axion DM is compelling and is especially attractive as it only works for a single value of the QCD axion mass.  As such, it is worth exploring all of the components that go into its calculation.  Surprisingly there are unresolved questions at almost every step of the way.

The scaling solution is itself a source of active research~\cite{Gorghetto:2018myk,Buschmann:2019icd,Gorghetto:2020qws,Buschmann:2021sdq}.  At this point, it is becoming clear that there exists a scaling solution where the energy density in the string network is logarithmically increasing in time.  While this overall structure and the length distribution of string is converging, we do not have a completely satisfactory analytic understanding of this behavior.  It would be interesting to obtain a more theoretical understanding of these distributions.  What is currently of particular interest about the scaling solution is energy spectrum of emitted axions.  There has been much theoretical and numerical work on this subject but it is currently unresolved, see e.g. Ref.~\cite{Dine:2020pds} for a recent attempt.  The spectrum of emitted axions is important because the number density of emitted axions is the string contribution to axion DM.  The current results vary from saying that the string contribution is dominant to completely irrelevant.  It would be interesting if more light could be shed on this debate.

Another aspect of the string network that needs to be explored carefully is the affect of friction on the string network.  QCD axion strings have large geometric scattering cross sections with gluons and thus the thermal plasma provides an important source of friction for axion strings~\cite{Agrawal:2020euj}.  When the scaling solution is found numerically, the thermal bath is ignored and thus these important friction terms are neglected.  It can be shown that at high temperatures, the gluon sourced friction is extremely important, while at temperatures near the QCD phase transition it is likely negligible.  It would be interesting if this friction term had any observable consequences or changed the predicted axion number abundance.

Another area that requires more detailed analysis is how the axion potential turns on around the QCD phase transition~\cite{Borsanyi:2016ksw}.  There have been lattice computations that give the axion mass around the minimum as a function of temperature.  However, there is no understanding of how the axion potential behaves away from 0 as a function of temperature.  This is important because the axion potential at high temperatures is well approximated by instantons and is in the form of a cosine.  At low temperatures, the axion potential is very much not a cosine and is rather cuspy at $\bar{\theta} = \pi$.  The behavior of the axion potential near $\bar{\theta} = \pi$ is important because if the axion field value started off near $\bar{\theta} = \pi$, then interesting dynamics such as axion miniclusters occur and may be interesting observationally.  Because the axion expectation value is equally distributed, the behavior at large $\bar{\theta}$ is necessarily important and should be explored.

\section{Axions and thermal friction }
\label{sec:thermalinflation}

Coupling an axion to a dark $SU(N)$ gauge group, a natural pairing, gives rise to thermal friction through sphaleron processes.
Thermal friction has intriguing phenomenological consequences in cosmological contexts when acting upon a scalar field that forms a dark energy-like component.
The idea is that a (pseudo-)scalar field $a$ couples to light degrees of freedom, such that a macroscopic friction coefficient $\Upsilon$ emerges.
The macroscopic friction acts on the equation of motion of the scalar field as additional friction to Hubble friction $H$.
Furthermore, the interaction between the scalar field and the light degrees of freedom sources dark radiation, such that a steady-state-temperature ($T > H$) can be maintained even in an inflating universe. The equations that govern the time evolution of the scalar field and the radiation are given by: 
\begin{eqnarray} 
    \ddot{a} +\left(3H + \Upsilon \right) \dot{a} +V'(a) &=& 0 \, , 
    \nonumber \\ 
    \dot{\rho}_{\text{dr}} +4H \rho_{\text{dr}}  &=& \Upsilon\dot{a}^2   \, ,
    \label{eq:eqs_of_motion}
\end{eqnarray}
where $\rho_{\text{dr}}$ is the energy density of dark radiation and $V(a)$ is the potential of $a$.
This warm inflation scenario~\cite{Berera:1995ie, Berera:1995wh, Berera:1999ws, Berera:1998px, Berera:2008ar, Bastero-Gil:2016qru, Berghaus:2019whh}, has both interesting predictions for observations as well as theoretical upsides, such as avoiding Trans-Planckian field excursions \cite{ Kamali:2021ugx, Berera:2020dvn}.

A crucial ingredient of a successful warm inflation model is a microscopic theory that leads to a vanishing thermal back-reaction to avoid modifications to $V'(a)$, as well as an efficient macroscopic friction coefficient $\Upsilon$ in order to mantain a sizeable temperature ($T > H$) \cite{Yokoyama:1998ju}. 
A recent breakthrough has been the realization that an axion coupled to a non-Abelian gauge group:
\begin{equation} \label{eq:lagrangian}
\mathcal{L} \propto \frac{\alpha}{8\pi} \frac{a}{f_a} G^a_{\mu \nu} \tilde{G}^{a\mu\nu}  
\end{equation}
is the ideal candidate to provide a compelling minimal model that is able to satisfy those requirements \cite{Visinelli:2011jy, Kamali:2019ppi, Berghaus:2019whh}. 

Due to its shift symmetry, the axion is protected from thermal mass corrections, which suppresses the back-reaction. Furthermore, coupling the axion to an $SU(N)$ gauge group exploits sphaleron processes that lead to friction coefficient $\Upsilon \sim \left(N \alpha \right)^5 \frac{T^3}{f_a^2}$ \cite{Laine:2016hma, Berghaus:2019whh}, a quantity that can be derived from the sphaleron rate in $SU(N)$ \cite{Moore:2010jd}, where $\Gamma_{\text{sph}} = 2f_a^2 T \Upsilon$. Sphalerons are non-trivial field configurations  that provide a barrier between vacuua of the theory with differing topological Chern-Simons number $N_{CS}$. A nonzero speed of the axion $ \langle \dot{a} \rangle \neq 0$ due to an external potential $V(a)$ biases transitions between differing vacuaa in one direction, thus generating sphalerons \cite{Berghaus:2020ekh}. The non-perturbative phenomena makes the axion a prime candidate for a warm inflaton.

This direction opens up many interesting research questions. A particularly compelling prospect is a model in which the QCD axion could be the inflaton, and QCD makes up the radiation that is sourced during inflation. This could allow for a scenario in which one smoothly transitions from an inflating universe to one that is Standard Model radiation dominated with minimal BSM requirements. 
Unfortunately the presence of light fermions in QCD generate a chemical potential that counteracts the bias giving rise to sphaleron transitions \cite{Berghaus:2020ekh}, not allowing for a sustained period of accelerated expansion in the early universe. A mechanism that gives larger masses to light quarks in the early universe could potentially avoid these constraints.     

Another interesting connection is the one to Lepto- and Baryogenesis. Sphaleron transitions violate baryon and lepton numbers, a property that makes them a popular mechanism to transmute lepton asymmetry to a baryon asymmetry in the Standard Model \cite{Buchmuller:2005eh}. Biased sphaleron transitions are built into the thermal axion friction mechanism, a fact that may be exploitable in extensions that also aim to address the matter-antimatter asymmetry.    

More work is required to understand the scaling of the macroscopic friction coefficient $\Upsilon$ in various regimes. Using the dissipation-fluctuation theorem, $\Upsilon$ can be derived from the sphaleron rate in the limit in which it amounts to a transport coefficient \cite{Laine:2016hma}. This condition is satisfied when the mass of the axion is much smaller than the interaction rate of the radiation which scales as $\sim \alpha^2 T$. In the strong regime of warm inflation ($\Upsilon > H$), this condition is easily satisifed. However, approaching the weak regime of warm inflation ($\Upsilon \lesssim H$) saturates the limit, potentially modifying the scaling of $\Upsilon$. A better understanding of this would be very interesting especially as extrapolation suggests that there is a regime around $\Upsilon \sim H$ in which a minimal model with a simple quadratic field potential is able to match the inflationary observations in the tensor-to-scalar ratio $r$ vs. spectral index $n_s$-plane. 

Another regime that is not fully understood yet is the one in which the coupling constant $\alpha$ approaches the strong limit. This is a difficult regime to calculate and lattice calculation of the sphaleron rate for $\alpha > 0.1$ is an active on-going area of research \cite{Altenkort:2020axj}. 

Interesting interplay exists between friction caused by tachyonic instabilities \cite{Anber:2009ua, Domcke:2019lxq} and thermal friction \cite{Ferreira:2017lnd, Ferreira:2017wlx}. Both Abelian and non-Abelian gauge groups coupling to an axion can develop tachyonic instabilities that populate a non-thermal background of very low frequency vectors ($\sim \frac{\dot{a}}{f_a}$). This effect arises due to an instability on the mode equation for the vector. The instability shuts off if the background modes thermalize. One mechanism can precede the other, setting the initial conditions for thermal friction to take over \cite{DeRocco:2021rzv}. The nonlinear friction arising from tachyonic instability has a different set of evolution equations, that requires tracking each individual mode which can be computationally demanding \cite{Domcke:2020zez}. 
Both mechanisms provide a tool that can extract energy from a scalar field and convert it into radiation. 

This is relevant not only for inflationary models but also for other cosmological epochs that make use of rolling (pseudo)scalar fields, such 
as late universe dark energy \cite{Graham:2019bfu, DallAgata:2019yrr, Berghaus:2020ekh}, as well as other models that address a variety of problems ranging from fine-tuning of the Higgs mass \cite{Hook:2016mqo} to a solution to the Hubble tension \cite{Berghaus:2019cls}. In those contexts, axion friction has proven itself to be a phenomenon that leads to novel predictions for observables while being a valuable model building tool. 
There may be many additional interesting applications that have not yet been explored.   


\bibliographystyle{JHEP}
\bibliography{ref}








\end{document}